\documentclass[prd,aps,showpacs,superscriptaddress,twocolumn]{revtex4}

\usepackage{graphicx}

\begin{document}

\title{
Properties of light scalar mesons from lattice QCD.}

\author{C. \surname{McNeile}}
\author{C. \surname{Michael}}

\affiliation{Theoretical Physics Division, Dept. Math. Sci., University
of Liverpool,  Liverpool L69 7ZL, UK.}

\collaboration{UKQCD Collaboration} 
\noaffiliation

 \begin{abstract} 
 Lattice QCD with $N_f=2$ flavours of sea quark is used to explore the
spectrum and decay  of scalar mesons. We are able to determine  the
$b_1$ - $a_0$ mass difference and this leads to the conclusion
that the  lightest non-singlet scalar meson ($a_0$) has a mass of 
1.01(4) GeV. We determine from the lattice the  coupling strength  to KK
and $\pi \eta$. We compute the leptonic decay constant of the lightest
non-singlet scalar meson.
 We discuss the impact of these lattice results on the interpretation of
the  $a_0(980)$ state. We also discuss $K^*_0$ states.
 \end{abstract}

\pacs{12.38.Gc, 14.40.Cs, 13.25.-k}

\maketitle

\section{Introduction}

The scalar mesons known experimentally do not fit into a tidy pattern, 
as found for vector or axial mesons, for example.  Because the scalar 
mesons have S-wave decays to light two-body states (two pseudoscalar
mesons),  then the impact of these two-body channels on the scalar meson
can be sizeable. Thus the scalar mesons may have $\bar{q}\bar{q}qq$ as
well as  $\bar{q}q$ components. For example, there are two $a_0$ 
mesons, $a_0(980)$ and $a_0(1450)$, known~\cite{Eidelman:2004wy}. The
$a_0(980)$ meson is closely associated  with the $\bar{K}K$ threshold
and it has been suggested that this is a  molecular state. This is can
be explored using lattice techniques.  A further complication is that
the flavour  singlet scalar mesons can mix with scalar glueballs,
although here we restrict our investigation to  the flavour non-singlet
scalar mesons from lattice QCD.

There has been a long history of studying the scalar non-singlet mesons
on the lattice. These states tend to have a poorer signal to noise ratio
than the S-wave mesons~\cite{DeGrand:1992yx} such as the $\rho$ and
$\pi$, hence are less commonly studied. Much of the early literature on
light P-wave mesons focussed on designing good interpolating operators to
create the  mesons~\cite{DeGrand:1992yx,Lacock:1996vy}.

The quenched studies of the $a_0$ were complicated by the  discovery of
a ghost state that  made  the correlator for the $a_0$ particle, which
should be  positive definite in a unitary quantum field theory, go
negative~\cite{Bardeen:2001jm}. If this effect was not taken into
account then the chiral extrapolation of correlators was unreliable.
Modern studies of this state such as those by Burch et
al.~\cite{Burch:2006dg} correct for  the effect of the missing
contribution to the $0^{++}$ correlator from the $\eta'$ meson.
Prelovsek~\cite{Prelovsek:2005rf} has also studied the ghost state in
the  $a_0$ correlator using 2+1 dynamical staggered fermions and mixed
(chiral valence and staggered sea) fermions. In both cases, which are
essentially partially quenched, deviant features are discovered.

The non-singlet scalar mass is an input into the study of mixing with
glueballs in the  singlet sector by  Weingarten and
Lee~\cite{Lee:1999kv}.  The ghost state was not taken into account and
this led to problems with the chiral extrapolation of the non-singlet
$0^{++}$ meson masses. This mixing has also been  
discussed~\cite{McNeile:2000xx,Hart:2002sp} using unquenched lattices 
which avoids this problem.

Alford and Jaffe~\cite{Alford:2000mm} used quenched QCD with
$\overline{q}^2 q^2$ operators relevant to  $0^{++}$ mesons. Their study
claimed to see evidence for bound states in the $\overline{q}^2 q^2$
channel relevant to $0^{++}$ states. The work  of Alford and
Jaffe~\cite{Alford:2000mm} can be criticised for not taking into account
the quenched ghost in the $a_0$ correlator. Only a subset of the
correlators required for the  singlet channel were computed. This is,
perhaps, consistent in quenched QCD but clearly important physics is
omitted.

The scalar collaboration are starting to use lattice QCD
techniques to study the $\kappa$ particle~\cite{Kunihiro:2005vw}.

 Prelovsek et al.~\cite{Prelovsek:2004jp} extended the work of Bardeen
et al.~\cite{Bardeen:2001jm} on the  effect of the ghost state in the
$a_0$ channel to  the partially quenched theory. By restricting lattice
study to valence quarks heavier than the  sea-quarks, Hart et
al.~\cite{Hart:2002sp} were able to extrapolate to  light quarks with no
ghost contributions, obtaining  an estimate for the $a_0$ mass of 
1.0(2) GeV.

In table~\ref{tab:a0quenched} we collect together some recent numbers
for the mass of the $a_0$ mass from some modern lattice calculations
that take into account the ghost term. None of the calculations in
table~\ref{tab:a0quenched} had complete control over all systematic
errors, such as finite size effects or the continuum limit,  even within
quenched QCD. The results for the lightest $0^{++}$ meson  are mostly
around 1.5 GeV.  As we note above, the $a_0$  decays via the strong
interaction, so a quenched QCD calculation may give a poor estimate of
the particle mass.

\begin{table}[tb]
\begin{tabular}{c|c|c}
Group & Method &  $m_{a_0}$  GeV \\ \hline
Bardeen at al.~\cite{Bardeen:2001jm} & quenched & $1.34(9)$ \\
Hart et al.~\cite{Hart:2002sp}  & $n_f=2$, partially quenched, 
                                     & $1.0(2)$ \\
Prelovsek et al.~\cite{Prelovsek:2004jp}  & $n_f=2$, unquenched, 
                                     & $1.58(34)$ \\
Prelovsek et al.~\cite{Prelovsek:2004jp}  & partially quenched
                                     & $1.51(19)$ \\
Burch et al.~\cite{Burch:2006dg} & quenched & $\sim 1.45$ \\
\hline
  \end{tabular}
  \caption{
 Some results for the mass of the $a_0$ meson from  quenched and
partially quenched QCD that include the effect of the  ghost
state~\cite{Bardeen:2001jm}.
}
\label{tab:a0quenched}
\end{table}

 The MILC collaboration reported evidence for  $a_0$ decay on the
lattice in an unquenched lattice QCD calculation with $2+1$ flavours of
improved staggered fermions with a lattice spacing of 0.12
fm~\cite{Bernard:2001av}. In MILC's first paper they found  the
$a_{0}$ mass to be significantly lower than the mass of the $b_1$ and
$a_1$ mesons. This was different behaviour from the quenched study with
the same parameters.
 MILC~\cite{Bernard:2001av}  found that, for lighter quarks,  the mass
of the $a_0$ meson was close to the sum of the  $\pi$ and $\eta$ masses, 
where the mass of the $\eta$ was estimated using the Gell-Mann-Okubo
formula.
 As the MILC collaboration~\cite{Aubin:2004wf} ran unquenched
calculations with even lighter sea quarks they confirmed that the
lightest state in the  $a_{0}$ channel lay below the $\pi\eta$
threshold.  Using independent techniques on a subset of the
configurations from MILC, Gregory et al.~\cite{Gregory:2005yr}  also
found that the lightest  state in the $a_0$ channel was below the
$\pi\eta$ threshold. Prelovsek~\cite{Prelovsek:2005rf} has studied 
$a_0$ decay using staggered chiral perturbation theory, concluding that 
taste violations in the staggered  fermion formalism  allow a small
amplitude for the  decay of the $a_0$ state to two pions. The decay $a_0
\rightarrow \pi \pi$ is forbidden in the real world because of $G$
parity.

 Since the current state of lattice investigation of scalar mesons is 
incomplete, more work is needed. 
 In order to make a start in establishing the nature of scalar mesons 
from first principle in QCD, we address here the flavour non-singlet 
scalar mesons. As is well known, lattice QCD in the quenched
approximation  is not a consistent theory and this manifests itself as
ghost contributions to  the scalar meson propagation - arising from the
spurious low-lying threshold in the  $\pi \eta$ two-body channel. 
 We use here $N_f=2$ dynamical gauge configurations so that we have  a
consistent field theory. The physical case, however, also has another 
light quark (the $s$ quark) and has lighter $u$ and $d$ quarks than we
are able to  use on a lattice. Thus some extrapolation will be needed to
obtain consequences for the physical spectrum. 

As we approach the limit of physical light quark masses, the scalar
mesons  become unstable: they are resonances. On dynamical lattices
these  decay channels are open. Thus we need to have methods to cope 
with unstable particles on a lattice.   The study of hadronic decays
from the lattice is not straightforward -  see
ref.~\cite{Michael:2005kw}. It is possible, however, to evaluate the
 appropriate hadronic matrix element from a lattice if the transition is
 approximately on-shell. This allows us to estimate decay widths, 
provided that the underlying coupling is relatively insensitive to 
the quark masses. 
 We follow methods  generically similar to those used  by us to study
$\rho$ decay~\cite{McNeile:2002fh} and hybrid meson 
decay~\cite{McNeile:2006bz}. 
 
As well as the hadronic decay, one can also define a {\em decay
constant}  analogously to that defined for the weak decay of
pseudoscalar mesons. We discuss the  relevance of this and the
determination from the lattice of the scalar decay constant.

\section{Spectrum}

As a by-product of our study of hybrid mesons, we have accurate lattice
measurements  of the $a_0$, $b_1$ and $a_1$ mesons from  clover-improved
lattices with $N_f=2$ degenerate sea-quarks - see Table~\ref{tablat} and
Table~\ref{tablatp} for details.  Each of these mesons is unstable and
in the $N_f=2$ world with two degenerate  quarks they have two-body
decays to $\pi \eta_2$, $\pi \omega$ and $\pi \rho$ respectively. Here
$\eta_2$ is flavour singlet, $(\bar{u}u +\bar{d}d)/\sqrt{2}$, so it is
more  like the $\eta'$ than the $\eta$ meson. Indeed
estimates~\cite{Allton:2004qq} of its mass from a mixture of lattice
results  and experiment suggest that it is near 0.86 GeV for light
quarks of physical mass. Thus, for these light quarks with $N_f=2$, the 
open decay channel is heavier for the $a_0$ meson than for the $a_1$ and
$b_1$  mesons.  Hence, in the self-consistent world with $N_f=2$ 
degenerate light quarks, we do not expect the $a_0$ meson to have any
peculiar features  compared to the other P-wave mesons.

 This is in contrast to {\em quenched} QCD where the flavour singlet
pseudoscalar  has the same mass as the pion, but an anomalous coupling.
Moreover,  quenched QCD allows a contribution (hairpin diagram) to the
$a_0$ correlator  from  this two-body  channel which gives significant
unphysical effects.

The conventional way to extract the mass of a meson is to use lattice
simulations at  successively smaller quark masses and to extrapolate
using an expression  based on chiral perturbation theory. For dynamical
simulations, which are mandatory here, one has a very limited range of
quark mass available. Resorting to partially quenched methods to  reach
lighter valence quarks is potentially dangerous, if the valence quarks
are lighter than the sea quarks. Indeed a  study using partially
quenched methods on the U355 and U350 data sets  has been
conducted~\cite{Hart:2002sp} and yielded an estimate of the $a_0$ mass 
of 1.0(2) GeV.  Here we explore a more reliable way to obtain the $a_0$
meson mass. 

Since the decay channels open to the P-wave mesons are quite similar, we
 propose to focus on the mass differences between them since this will 
reduce lattice artifacts. 
 The $a_1$ meson is very wide, experimentally,  so that it is not a good
point of comparison with lattice results. The $b_1$ meson, however  is
relatively narrow and should be well reproduced on a lattice. Indeed in 
ref.~\cite{McNeile:2006bz}, we were able to measure from the lattice the
decay amplitude  for the S-wave decay $b_1 \to \pi \omega$, obtaining
agreement with experiment.
 For lattice U355, for example,the decay threshold is at  0.72(4) for
$a_0$ from  $\pi \eta_2$ and at 0.883(8) for $b_1$ from $\pi \omega$, in
lattice units. These  energy values are both above the mass values we
report in Table~\ref{tablatp}, so each state is stable on our lattice
and they are about equally  below the lowest threshold.

We show our results from two state fits to a  $2\times3$ matrix of
correlators ($2\times2$ for U350 and C390) using $t$-range 3-12 (3-10
for C390 and C410) in Table~\ref{tablatp}. The methods used are 
described in more detail in ref.~\cite{McNeile:2006bz}. We use local 
and extended sources at the source (and two sizes of extension at the
sink in some cases). The excited mass values are in all cases
significantly higher  (by over 50\%) than the ground state values
reported and do not correspond to any simple two-body level. Thus the 
$\pi \eta_2$ threshold level at $aE=0.72$ for U355 does not feature
in the fit. As we shall see later, this is consistent with the
relatively weak transition amplitude on a lattice  between two-body
states and the $a_0$.

We find that the  $a_0$ correlator can have big fluctuations which are
apparent  at large $t$, most noticeably for U350 where the zero-momentum
effective  mass decreases at large $t$. The origin of these fluctuations
is  mixing between the $a_0$ and the  pion induced by regions of 
odd-parity  in the vacuum - presumably associated with instantons. 
See~\cite{Faccioli:2003qz} for a discussion of $a_0-\pi$ mixing in lattice
QCD and in the instanton liquid model. With sufficient statistics these
odd-parity fluctuations average to zero.  Using stochastic methods
(all-to-all) helps to reduce these fluctuations
 as we reported before~\cite{McNeile:2006bz}. Using non-zero momentum
can also act as a  useful cross-check.  This suggests that  for U350
with zero momentum,  we should use a $t$-range from 3-8 to reduce these 
fluctuation effects and retain consistency with our results from
momentum $2 \pi n/L$ where $n=(1,0,0)$ and $(1,1,0)$. The value for U350
quoted in Table~\ref{tablatp} is from  this analysis.

For the $b_1$ meson, at non-zero momentum there can be mixing  with the
$\rho$ meson (for some spin states). For the non-local (fuzzed)
operators  there will also be an admixture of  $L=2$ (from distortion
due to Lorentz boost)  and possibly some mixing of opposite C (unless 
the momentum phase factors are applied symmetrically to the fuzzed
operator).
 For these reasons we rely on  zero momentum for the $b_1$ meson.

\begin{figure}[t]
  \centering
    \includegraphics[width=.8\linewidth]      {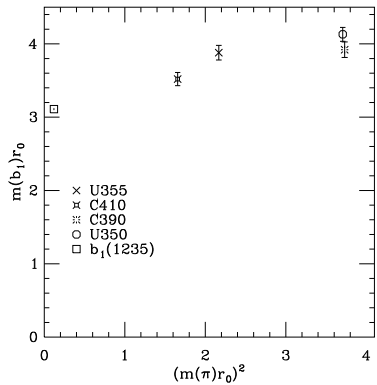}
        \caption{ Mass  of the $b_1$  mesons (in 
units of $r_0 \approx 0.5$fm) versus quark mass.  
The strange quark mass corresponds to $(m(\pi)r_0)^2 \approx 3.4$.
 }   
      \label{fig:mb1}
      \end{figure}

 Since the $b_1$ meson has an  unambiguous interpretation as
predominately a bound state of a quark and anti-quark, we show our 
spectrum results for it versus quark mass in fig~\ref{fig:mb1}. Here we
see  that our lattice results are quite consistent with a smooth
extrapolation  to the experimental mass value for physical light quarks.
To have a precision determination of the mass would require a continuum
extrapolation as well  as an extrpolation in quark mass and we do not
have data sufficient to  undertake this combined extrapolation.

\begin{figure}[t]
  \centering
    \includegraphics[width=.8\linewidth]      {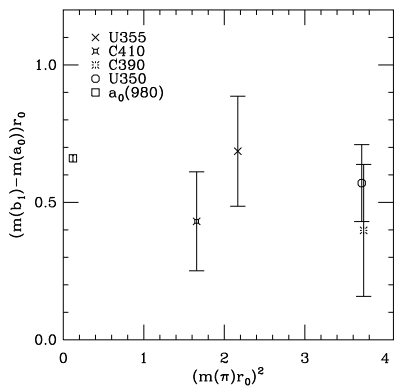}
        \caption{ Mass difference of $b_1$ and $a_0$ mesons (in 
units of $r_0 \approx 0.5$fm) versus quark mass.  
 }   
      \label{fig:mba}
      \end{figure}

Because of the difficulties in extrapolating to light quarks using
lattice results  with a range of different lattice spacings, we focus on
mass differences.  Here we concentrate on the difference $m(b_1)-m(a_0)$
which is plotted against the quark  mass in fig.~\ref{fig:mba}  using
$r_0$ determined on the lattices to create a dimensionless comparison.

\begin{table}

\begin{tabular}{lllllll}

Code &  no. & $\kappa$ & $m(\pi)r_0$ & $r_0/a $ & 
 $am(\pi)$ & $am(\rho)$ \\
\hline 

C410 & 237 & 0.1410 & 1.29 & 3.01 & 0.427(1) & 0.734(4) \\ 

C390 & 648 & 0.1390 & 1.93  & 2.65  & 0.729(1) & 0.969(2)   \\ 

U355 & 200 & 0.1355 & 1.47 & 5.04 & 0.292(2) & 0.491(7) \\ 

U350 & 151 & 0.1350 & 1.93 & 4.75 & 0.405(5) & 0.579(8)  \\

\end{tabular}

\caption{Lattice gauge configurations U355 and U350 from
UKQCD~\cite{Allton:2001sk} and C390 and C410 from
CP-PACS~\cite{AliKhan:2001tx} are used, all having spatial extent L=16a.
 These have $N_f=2$ flavours of sea quark  and we use valence quarks  of
the same mass as the sea quarks.
 } 
\label{tablat}
\end{table}

\begin{table}

\begin{tabular}{llll}

Code  &  $am(b_1)$& $am(a_1)$ & $am(a_0)$\\
\hline 

C410 &  1.17(3) &    1.15(2) & 1.03(4) \\ 

C390 & 1.48(4) &  1.39(5) & 1.33(8) \\ 

U355 & 0.77(2) &  0.72(2) & 0.64(4) \\

U350 &  0.87(2) & 0.88(2) & 0.75(3) \\ 

\end{tabular}

\caption{Results for P-wave mesons  from the methods of
ref.~\cite{McNeile:2006bz} for U355 and C410 and from conventional
methods for U350 (with 4 time sources)  and C390.
 } 
\label{tablatp}
\end{table}


The point in fig.~\ref{fig:mba} labelled $a_0(980)$ assumes that the 
relevant $a_0$ meson is the lightest with mass 984.7 MeV. The next
heaviest with mass 1474 MeV is less well established and would
correspond to a  point (-0.66) far below the $x$-axis. 
 Our lattice results for the mass difference show no significant
dependence on the quark mass, and averaging our lattice results gives an
estimate (using $r_0=0.5$fm) of  $m(b_1)-m(a_0)=221(40)$ MeV. There is
an  additional systematic error coming from the assumption of a constant
difference as the quark mass is decreased, which we are unable to
quantify.  As discussed above, our lattice masses in Table~\ref{tablatp}
 are at quark masses around the strange quark mass and at non-zero
lattice spacing.  They correspond, as expected,  to masses somewhat
larger than the physical $b_1$ mass of  1230 MeV, but are consistent
within the expected systematic errors of the extrapolations necessary.


As discussed above, we do not expect the two-body thresholds to play a 
significant role in our $N_f=2$ spectra. We do, however, measure these  
decay transitions to have a more complete analysis.

\section{Hadronic decays}

\begin{figure}[t]
  \centering
  \includegraphics[width=.8\linewidth]      {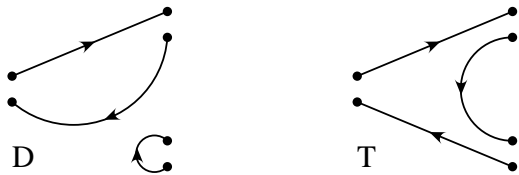}
        \caption{ Quark diagrams involved in the decays  listed in
Table~\ref{tabdt}, where  $D$ is the  disconnected diagram and $T$ is
the triangle diagram.
 }
      \label{fig:fd_a0}
      \end{figure}

For the case of $N_f=2$ degenerate quarks, the matrix elements for 
decay transitions of a non-singlet scalar meson to two pseudo-scalar
mesons are given in  Table~\ref{tabdt}, where the quark diagrams are
illustrated in fig.~\ref{fig:fd_a0}.

\begin{table}

\begin{tabular}{lllll}

  S     &  $P_1$ & $P_2$ & $T$ & $D$ \\ 
\hline 
$a_0 $  & $\pi$ & $\eta_2$    & $2^{1/2}$ & $-2^{1/2}$ \\

$a_0 $  & K     &$\bar{\mathrm{K}}$&       1      &            0 \\
$a_0 $  & $\pi$ & $\eta_{ss}$ &   0          & $   -1$     \\
$a_0 $  & $\pi$ & $\eta_{8}$  & $(2/3)^{1/2}$ &  $   0$     \\
K$^*$   & K$^+$ &  $\pi^0$    & $2^{-1/2}$ & 0           \\
K$^*$   & K$^0$ &  $\pi^+$    & $1$          & 0           \\
K$^*$   & K     &  $\eta_2$   & $2^{-1/2}$ & $-2^{-1/2}$ \\
K$^*$   & K     & $\eta_{ss}$ & $1$          & $-1$ \\

\end{tabular}

\caption{Coefficients of transition amplitudes  from flavour
non-singlet scalar meson S to $P_1 P_2$ for the
triangle  quark diagram ($T$) and the disconnected  quark diagram ($D$).
Only the top line is allowed if $N_f=2$ strictly.  The other lines are
allowed when a valence $s$ quark is added. We define $\eta_2$ as
$(\bar{u}u +\bar{d}d)/\sqrt{2}$, $\eta_{ss}$ as $\bar{s}s$  and $\eta_8$
as $(\bar{u}u +\bar{d}d-2\bar{s}s)/\sqrt{6}$. We have assumed that the
disconnected  contributions to the decay to $\eta_8$ cancel.
 } 
 \label{tabdt}
\end{table}

Only one case, $a_0 \to \pi \eta_2$, is allowed staying strictly within
$N_f=2$  with valence quarks of the same properties as sea quarks (here
$\eta_2$ is  the flavour singlet pseudoscalar for $N_f=2$, namely 
$(\bar{u}u +\bar{d}d)/\sqrt{2}$). This case involves a disconnected
diagram (D) and is not directly relevant to phenomenology.  
 In the limit that the strange quark is much heavier than the $u$ and $d$
quarks,  we expected the neglect of $s$ quarks in the sea to be a good
approximation. In that case,   decays such as $K_0^* \to K \pi$  
 can be studied from diagram $T$. For the physical case with $s$ quarks of
some 80 MeV, $a_0 \to \pi \eta_8$  and $a_0 \to \bar{K} K$  may also be
determined adequately  from $N_f=2$ lattice study of diagram $T$.

With this in mind, we first evaluate the lattice transition amplitude
corresponding to  the connected triangle diagram $T$. The contribution
of $T$  to various decay amplitudes will have the  numerical factors
listed in Table~\ref{tabdt}. The most relevant cases will be  $a_0 \to K
\bar{K}$ and $K^* \to K \pi$.  This is  a partially-quenched
evaluation in the sense that we use valence $s$-quarks (of the same mass
 as our $u$, $d$ sea-quarks) which are not present in the sea. We are
able to  use similar methods to those  used  to study $\rho$
decay~\cite{McNeile:2002fh} and hybrid meson decay~\cite{McNeile:2006bz}.

 The lattice results for the connected ($T$) contribution to a generic
scalar meson  transition to two pseudoscalar mesons are presented as the
normalised lattice ratio
  $$
 R(t) = { (S \to P_1 P_2 ) \over \sqrt{(S \to S) (P_1 \to P_1) (P_2 \to P_2)} }
 $$
 where the three-point correlator is constructed from propagators  as
illustrated for $T$ in fig.~\ref{fig:fd_a0}. Each two and three-point
correlator  is taken at the same time separation $t$.

 Since the $a_0$ mass is approximately twice the pseudoscalar mass (see
Table~\ref{tablatp})  at zero momentum, we have an on-shell transition
and we  expect~\cite{McNeile:2002fh,McNeile:2006bz} the ratio $R(t)$ to
be approximately linear  with slope $xa$ versus $t$ where $x$ is the
lattice transition amplitude. This is indeed observed, as shown  in
fig.~{fig:xta0}.

 We first checked that using different operators to create mesons gave
essentially  the same ratio $R(t)$. We use local or fuzzed operators 
for each of the three particles involved and in each case the ratio is 
the same within errors for the $t$ region of interest for the case we
studied  in most detail, namely with all momenta zero.

\begin{figure}[t]
  \centering
  \includegraphics[width=.8\linewidth]      {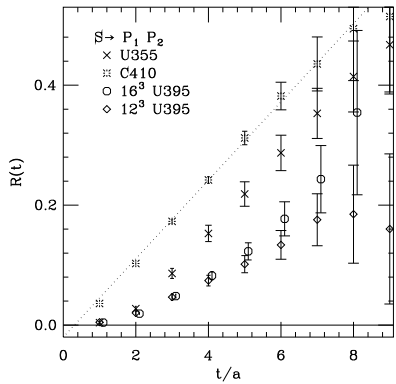}
        \caption{ The normalised ratio $R(t)$ for the connected
contribution ($T$) to the transition $S \to P_1 P_2$. The contribution
of $T$  to particular decays can be read off from Table~\ref{tabdt}. The
number of  lattice gauge configurations  analysed  was 90 (U355), 165
(C410)  and 30 for each U395 case.  The dotted line  illustrates the
expected behaviour with slope $xa$ for C410. 
 }
      \label{fig:xta0}
      \end{figure}

The most reliable determination of the coupling constant comes from using 
meson operators which minimise excited  state contributions. We use 
fuzzing with separations of $3a$ (C410) or $5a$ (U355) to achieve this. 
 We extract the slope $xa$ by taking finite differences and relate this
lattice transition amplitude to the continuum coupling via Fermi's
Golden Rule. The derivation of the phase space factor is described 
in ref.~\cite{statich}.
Then, to compare different lattice  data sets, we extract
the effective coupling using~\cite{McNeile:2002fh,CMcairns,McNeile:2006bz}
 $$
 {g}^2 = {1 \over  \pi} (xa)^2 (L/a)^3  { a E(P_1) E(P_2) \over E(P_1)+E(P_2)}
 $$
 Here  the decay width $\Gamma$ is, for a process with amplitude $T$,
given by  $\Gamma/k={g}^2$, where $k$ is the centre of mass momentum of
the decay products. For particular transitions, the quark coupling
coefficients of Table~\ref{tabdt} also enter, squared, in the decay rate.

 As a first check of this approach, we evaluated the effective coupling 
from lattices that differ only in spatial size (labelled U395, see
ref.~\cite{McNeile:2006bz} for  more details) and we found excellent
agreement when the spatial volume was changed by a factor of 2.4., as 
shown in fig.~\ref{fig:xta0}.  

\begin{figure}[t]
  \centering
  \includegraphics[width=.8\linewidth]      {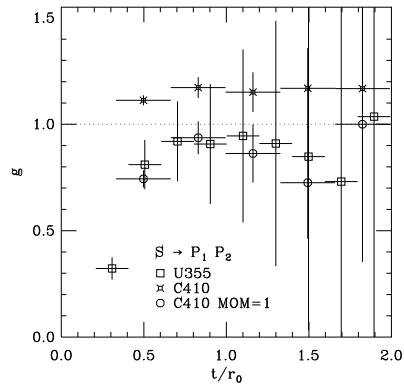}
        \caption{ The effective coupling $g$ extracted from $R(t)$ as
described in the text for the triangle graph $T$ for $S \to P_1 P_2$
with zero momentum (also some results for non-zero momentum as discussed
in the text). The dotted line at $g=1$ is to guide the eye.
 }
      \label{fig:a0f}
      \end{figure}

The coupling extracted, as above,  from our higher statistics data-sets
is shown in fig.~\ref{fig:a0f}. This shows a  coupling $g \approx1$
which has implications which we discuss later. The consistency  between
the two determinations (C410 and U355) which have different spatial 
volumes and different lattice spacings is satisfactory. As an overall
summary  we quote a coupling $g=1.0(2)$.

\begin{figure}[t]
  \centering
  \includegraphics[width=.8\linewidth]      {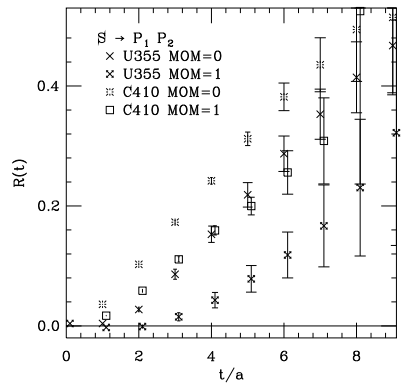}
        \caption{ The normalised ratio $R(t)$. Here MOM=0,\ 1 refers to the 
transition $S(k) \to P_1(0) + P_2(k)$ with momenta $k=2n \pi/L$  with
$n=0$ and 1.
 }
      \label{fig:xta0k1}
      \end{figure}

We also have available some results (from 40 gauge configurations for
U355  and for 50 for C410)  for transitions involving non-zero momentum,
especially $S(1) \to P_1(0) + P_2(1)$ where the momentum (in units of 
$2 \pi/L$) is given in the brackets. 
 These results used the methods  of
ref.~\cite{McNeile:2002fh,McNeile:2006bz}  respectively. The normalised
lattice ratio $R(t)$  is shown in fig.~\ref{fig:xta0k1} and the coupling
extracted  assuming the formulae above is included for C410 (where we
used an optimum  method to extract ground state contributions) in
fig.~\ref{fig:xta0}. As discussed in  ref.~\cite{McNeile:2002fh}, the
decay in flight poses some problems of  normalisation (since it is not
quite equivalent to a centre of mass decay with relative momentum
$\pi/L$), so must have a somewhat bigger systematic error to compensate.
Nevertheless, we see an approximate agreement of the  lattice transition
amplitude $xa$ and of the coupling $g$ when the decay  has  momentum
release  of zero and of  $\pi/L$. This is to be expected for  an S-wave
decay where the effective matrix element should be independent of
momentum.

\section{Leptonic decay constant}

The decay constants of non-singlet $0^{++}$ mesons are not routinely
calculated using lattice QCD, although they  are of  interest for a
number of  reasons. The value of the decay constant, which is basically
the  amplitude to find  a quark and anti-quark at the origin, can help 
distinguish between different quark content of the 
meson~\cite{Maltman:1999jn,Narison:2005wc}. For instance, if the $a_0$
was a $\overline{K}K$ molecule, then the decay constant would be small
relative to the value of the pion decay constant. The decay constant of
the $a_0$ meson is also a theoretical input to study of B meson decays
and of $\tau$ decays to final states that include an 
$a_0$~\cite{Diehl:2001xe,Aubert:2004hs,Couderc:2005cx,Delepine:2005px}.

Diehl and Hiller discuss the prospects of determining the  value of the
decay constant of the $a_0$  mesons from experiment~\cite{Diehl:2001xe}.
As we explain below, a direct measurement of the decay $a_0$ constant
coupled with computation of a QCD  matrix element could be used to
compute the mass difference of the up and down quarks.

 The  decay constant of the light $0^+$ meson can be defined by
equation~\ref{eq:VectorDefn}.
 \begin{equation}
\langle 0 \mid V_\mu^{ab} | a_0 \rangle
= i p_\mu g_{a_0}
\label{eq:VectorDefn}
 \end{equation}
where $V_\mu^{ab}$ is the vector current for quark flavours
$a$ and $b$.

The conservation of the vector current is used to  relate the operator
in equation~\ref{eq:VectorDefn} to the scalar current.
 \begin{equation}
\partial_\mu ( \overline{q^a} \gamma_\mu  q^b ) = i( m_a - m_b) \overline{q^a}
q^b
 \label{eq:vectorCurrent}
 \end{equation}
 for light quarks with flavour $a$ and $b$. This motivates
a definition of the decay constant such as
\begin{equation}
i \langle 0 \mid \overline{q^u} q^d \mid a_0 \rangle 
= \hat{f}_{a_0} m^2_{a_0}
\label{eq:anotherDEFN}
\end{equation}

To compare the size of the decay constant of the $a_0$ to that of the
$K^*(1430)$ meson, Maltman~\cite{Maltman:1999jn} defined a new decay
constant with a slightly different normalisation. 

The direct use of equation~\ref{eq:vectorCurrent} is impossible in a
lattice calculation with two degenerate flavours of sea quarks. The vector
current does not couple to the scalar meson in this case. The decay
constant in equation~\ref{eq:anotherDEFN} is non-zero in an unquenched
lattice QCD calculation with two flavours of sea quarks. 

There is another reason for splitting the definition of the 
$0^{++}$ meson decay constant into a quark mass factor and 
QCD matrix element.
Currently there is a disagreement between the value of the strange quark
from unquenched lattice QCD calculations that  use different
types of fermion for the light quarks~\cite{Rakow:2004vj}. 
Some recent
papers~\cite{DellaMorte:2005kg,Gockeler:2006jt,Mason:2005bj} 
report summaries of 
the values for the strange quark mass published around time of 
 the lattice 2005 conference,
using different formulations of lattice QCD.
Lattice QCD calculations are only just starting to report
values for the differences between the masses of the 
up and down quarks~\cite{Aubin:2004fs}.
Hence, we prefer to quote separately our measured 
matrix element rather than introduce explicit factors of the quark mass.
So, it is more natural to
define the decay constant using
equation~\ref{eq:decayDEFN}.  
 \begin{equation}
\langle 0 \mid \overline{q} q | a_0 \rangle
= m_{a_0} f_{a_0}
\label{eq:decayDEFN}
 \end{equation}
The relation between $f_{a_0}$ and $g_{a_0}$ is
via
\begin{equation}
g_{a_0} = \frac{m_d - m_u} {m_{a_0}} f_{a_0} 
\label{eq:connect}
\end{equation}
The explicit factor of the quark masses $(m_d - m_u)$ in 
equation~\ref{eq:connect} is the reason that 
Narison~\cite{Narison:1988xi} computes the value of 
$g_{a_0}$ to be between 1.3 and 1.6 MeV.

The matrix element in equation~\ref{eq:decayDEFN} is extracted
from the amplitudes in the fits 
to the correlators (see~\cite{Herdoiza:2006qv} for example).
The raw numbers from the lattice calculation need renormalisation. To
convert the lattice number to the $\overline{MS}$  scheme we use 
tadpole improved perturbation theory to one loop order~\cite{Lepage:1992xa}.
The
renormalisation factor for a scalar current, at the
scale $\mu=1/a$,
 is
 \begin{equation}
Z_S(\mu=1/a) = u_0  
\left( 
1 - \alpha_s S_c 
\right)
\label{eq:ZStad}
 \end{equation}
 where $u_0$ is the fourth root of the plaquette,
and the constant $S_c$ is $1.002$ for the Wilson gauge action~\cite{Bhattacharya:2000pn}
and
$0.5031$ for the Iwasaki gauge action~\cite{Taniguchi:1998pf,Aoki:1998ar}.

To remove $O(a)$ terms we also need to use improvement coefficients. We
define the renormalisation $\hat{Z}_S$ that includes the improvement
factor
 \begin{equation}
\hat{Z}_S = Z_S \left( 1 + b_S m_q \right)
\end{equation}
 where $m_q$ is the mass of the light quark. We used the one loop
expression for $b_S$.
 \begin{equation}
b_S = \left( 1 + \alpha_s b_{sc}  \right) 
 \end{equation}
where the constant $b_{sc}$ is $1.3722$ for the 
Wilson action~\cite{Sint:1997dj},
and $1.2800$ for the Iwasaki action~\cite{Aoki:1998qd,Aoki:1998ar}.
 We used the coupling computed in the $\overline{MS}$ scheme. For the
UKQCD data set we  used the coupling determined on the same data 
set~\cite{Booth:2001qp}. For the CP-PACS data we used the
$\overline{MS}$ coupling quoted in their paper~\cite{AliKhan:2001tx}.
The coupling was evaluated at the scale $\mu=1/a$.
Our results are in table~\ref{tab:OurResults}. 
As we only have decay constants for two different quark masses
with the same action,
we do not attempt a chiral extrapolation. The dependence
of the decay constant on the pion mass seems small, however.
In
table~\ref{tab:a0decother} we compare our results to other
determinations of the decay  constants.  The results in
table~\ref{tab:OurResults} show that the decay constant
$f_{a_0}$ is not suppressed relative to the pion decay
constant.

\begin{table}[tb]
\begin{tabular}{ccc|c}
$\kappa$ &  a $f_{a_0}/\hat{Z}_S$  &  $\hat{Z}_S$ & 
$f_{a_0}$  MeV \\ \hline
.1355 &  0.352(19)  &  0.70 &      488(26) \\
.1350 &  0.346(30)  &  0.71 &      460(40) \\
.1410 &  0.474(48)  &  0.79 &      478(48) \\ 
.1390 &  0.480(97)  &  0.84  &     513(104) \\
\hline
  \end{tabular}
  \caption{
Our results for decay constant of the $a_0$ meson.
}
\label{tab:OurResults}
\end{table}

The decay constant of the $0^{++}$ meson  is one of the parameters in
the model that gets rid of the ghost state in the  scalar $0^{++}$
correlator in  quenched QCD~\cite{Bardeen:2001jm} and partially-quenched
 QCD~\cite{Prelovsek:2004jp}, so there are estimates  for it. These
studies of the $0^{++}$ used another normalisation convention for the
scalar decay constant, so we do not tabulate their values here.

Chernyak~\cite{Chernyak:2001hs} uses a fit to data with
a factorisation assumption to obtain $g_{0^+}$ = $70 \pm 10$ MeV for the
$K^*(1430)$. Converting to our normalisation conventions, 
using a nominal value of the strange quark mass of 100 MeV,
this
corresponds to $f_{K^*(1430)}$ = $1000 \pm 140$ MeV. The results for the 
decay constants in table~\ref{tab:OurResults} are larger than
the results of UKQCD's recent calculation of the decay
constant of the $0^+$ charm-light meson~\cite{Herdoiza:2006qv}.

\begin{table}[tb]
\begin{tabular}{c|c|c}
Group & Method &  $f_{a_0}$  MeV \\ \hline
Maltman~\cite{Maltman:1999jn} & sum rule 
& $298$ \\
Shakin and Wang.~\cite{Shakin:2001sz}  & model 
                                     & $433$ \\
Narison.~\cite{Narison:1988xi}  & sum rule
                                     & $320-390$ \\ \hline

  \end{tabular}
  \caption{
Some results for decay constant  of the $a_0$ meson. We used
a value of $m_d - m_u$ of 4 MeV to convert the normalisation
of Narison's estimate. The quark masses quoted by 
Shakin and Wang were used to convert normalisation conventions
for other two results.
}
\label{tab:a0decother}
\end{table}

As an aside we note that if the decay constant $g_{0+}$
of the $a_0$ or $K^\star(1430)$ was measured experimentally,
then it would allow an additional method to measure
the quark mass differences $m_u - m_d$,  $m_s - m_d$
respectively, using lattice estimates of the QCD matrix elements.

\section{Discussion}

For the non-strange flavour non-singlet scalar meson ($a_0$), our 
lattice determinations using the self-consistent $N_f=2$ approximation
to QCD give  clear support for a physical $a_0$ meson lying
substantially lighter than the $b_1$. The mass estimates we find are
consistent with the observed  $a_0$ at 950 MeV but not the heavier
state at  1474 MeV. 

 To relate our approach to experiment with an additional strange quark, 
we can assume that the strange quark pair production is  relatively
small and so can be neglected. For K$_0^*$ propagation, for example, 
this amounts to treating the K$\pi$ channel correctly but having an
anomalous  contribution from K$\eta_{ss}$ intermediate states. Here the 
$\eta_{ss}$ propagation has a missing piece (just as $\eta_2$ does  in
the $\eta_2 \pi$ contribution to the $a_0$ propagation in quenched QCD
with $N_f=2$) and so will not have a single exponential but two
contributions with masses corresponding to (i) the  connected
pseudoscalar meson with valence quarks of strange mass and (ii) the
$\eta_2$ meson. Both of these contributions are  not especially light,
so we do not expect any major distortion of the  K$_0^*$ from using
valence $s$-quarks. Similarly the $a_0$ decays  to $K\bar{K}$ and
$\eta_8 \pi$ are expected to be accessible without major distortion from
the neglect of  strange quarks in the sea, as we discussed above.


 For the strange scalar mesons, the K$_0^*(1430)$ with mass 1412 MeV is
heavier than the  corresponding axial mesons (K$_1^*$ with masses 1273
and 1402 MeV). These two axial mesons are related to a mixture of the 
strange partners of the non-strange $b_1$ and $a_1$ mesons since charge
conjugation is not a good quantum number  for strange mesons. So the
interpretation in this case is  unclear. As well as this strange scalar
meson at 1412 MeV,  one might expect a lighter state, about 100-130 MeV 
heavier (mass split determined from tensor mesons) than the $a_0(980)$. 
The so-called kappa ($\kappa$) at 700-900 MeV with a very broad width 
(400 MeV or more) has been claimed by many
sources~\cite{Eidelman:2004wy} and a recent analysis~\cite{Bugg:2005xx}
gives mass $750^{+30}_{-50}$ MeV. There is no consensus yet on the
existence of the  kappa, because some analyses of experimental data see
no sign of it~\cite{Cherry:2000ut}.
 Our lattice studies  suggest that a scalar K$_0^*$ meson of mass around
1000-1200 MeV would  be expected in a theory with $N_f=2$ sea quarks and
a strange  quark treated as a valence quark.  For our case where the
valence $s$-quark and $u,d$ sea-quarks  have the same mass, the
anomalous $K \eta_{ss}$ intermediate state combines with the  $K \pi$
intermediate state to give only a $K \eta_2$ intermediate state, just
like  the case of  $a_0$ propagation. Hence our  lattice  treatment does
not correctly include the $K \pi$ threshold in the K$_0^*$ meson
propagation and so may be less reliable than the $a_0$ propagation.

The connected decay diagram ($T$) is appropriate for the decays 
$a_0 \to K K$, $a_0 \to \pi \eta_8$ and $K_0 \to K \pi$ where the 
appropriate factors are given in Table~\ref{tabdt}. 
 Then the experimental data~\cite{Eidelman:2004wy}  can be used to
estimate the coupling (from $\Gamma/k$). For K$_0^*(1430)$, this gives
$g^2= 0.32(3)$. While for the $\kappa$, one recent
analysis~\cite{Bugg:2005xx}  finds a width of  $342\pm60$\ MeV which
corresponds  to $g^2=0.7(2)$. 
 For $a_0(980)$,  the state is close to the $\bar{K}K$ threshold which
distorts the appearance of the  meson. Phenomenological analyses vary
but one example quotes~\cite{Achasov:2002ir} a total width of  153 MeV
and a coupling given by $g^2= 0.82$ for $\bar{K}K$ and  around 0.7 for
$\eta \pi$. 
 For $a_0(1450)$, the partial widths are not well known and one can only
estimate that the $\bar{K}K$ and $\eta \pi$ decays yield couplings
smaller than $g^2= 0.23$ and 0.34 respectively. 

Our determination of the coupling which controls decay is also relevant 
for identification of states.
 We find a coupling (normalised to diagram $T$ above) given by $g
\approx 1$. This favours  the lighter $a_0$ and $\kappa$ meson over the
heavier states. Our determination of the $a_0$ decay constant 
disfavours a molecular structure for this state, in agreement with  our
conclusion from hadronic decays. The only concern is that for the
$K^*_0$  meson the experimental evidence gives a $\kappa$ meson lighter
than our expectations.

\section{Conclusion}
 
We have studied the spectrum and decay of non-singlet scalar mesons from
first principles using  lattices with a consistent (unitary) field
theoretic  interpretation for $N_f=2$ flavours of sea-quark. Rather than
extrapolate the scalar masses directly, we concentrate on the
 mass splitting between the $a_0$ and $b_1$ mesons from the lattice. The
lattice results are unambiguous and point to a scalar meson which is
221(40) MeV lighter than the  $b_1$. Since the  experimental mass value
of the $b_1$ meson is 1230 MeV, this suggests that the  $a_0(980)$ is
indeed the lightest non-singlet scalar meson in a theory with $N_f=2$
flavours  of degenerate quark. Our approach does not  include the
$K\bar{K}$ channel, so this channel is to be regarded as having an
impact on  a pre-existing state, rather than as being the dominant
component of the state.  In other words, we do not find that a
$K\bar{K}$ molecule is a good approximation to the  $a_0(980)$.

Our results for the decay transition amplitude are also consistent with
the  phenomenological estimates of the coupling  of the $a_0(980)$ to
$K\bar{K}$  and $\eta \pi$. Overall, we conclude that the $a_0(980)$ is
predominantly  a conventional meson with normal couplings to $\bar{q}
q$.

For the $K_0^*$ scalar meson, we expect a mass 100-130 MeV heavier than
the $a_0$ (based, for example, on the observed mass splittings of the
tensor mesons). This is not easily related to any experimental
candidate: the  $\kappa$ is too light (700-900 MeV), while the 
$K_0^*(1430)$ is too heavy. What may help  clarification  is that we
find a decay coupling transition (to $K \pi$) which  is comparable to
that needed phenomenologically for the $\kappa$ but much larger than
that needed for the $K_0^*(1430)$. This suggests that the $\kappa$ is
more closely related to  the state obtained in $N_f=2$ lattice QCD.
A lattice treatment with the strange quark included in the sea would 
help to clarify further this conclusion.

\section{Acknowledgements}
 
The authors acknowledge support from  PPARC grant PPA/Y/S/2003/00176.
 This work has been supported in part by the EU Integrated
 Infrastructure Initiative Hadron Physics (I3HP) under contract
  RII3-CT-2004-506078.
 We acknowledge the ULGrid project of the University of Liverpool for making
 available  computer resources.
 We acknowledge the CP-PACS collaboration~\cite{AliKhan:2001tx} for
making available  their gauge  configurations. 

\newpage

\end{document}